\documentclass[prd,nofootinbib,twocolumn,showpacs]{revtex4}
\usepackage{graphics} 
\usepackage{bm} 
 
\begin{document} 
 
\title{Spontaneous formation of domain wall lattices 
       in two spatial dimensions} 
 
\author{Nuno D. Antunes} 
\affiliation{ 
Center for Theoretical Physics, University of Sussex, \\ 
Falmer, Brighton BN1 9WJ, United Kingdom.} 
 
\author{Tanmay Vachaspati} 
\affiliation{
CERCA, Department of Physics, Case Western Reserve University, 
10900 Euclid Avenue, Cleveland, OH 44106-7079, USA.} 

\begin{abstract} 
We show that the process of spontaneous symmetry breaking can 
trap a field theoretic system in a highly non-trivial state 
containing a lattice of domain walls. In one large compact space
dimension, a lattice is inevitably formed. In two dimensions, 
the probability of lattice formation depends on the ratio of sizes 
$L_x$, $L_y$ of the spatial dimensions. We find that a lattice 
can form even if $R\equiv L_y/L_x$ is of order unity. We numerically
determine the number of walls in the lattice as a function of 
$L_x$ and $L_y$.
\end{abstract}

\pacs{98.80.Cq, 05.70.Fh}

\

\maketitle

\section{Introduction}
\label{intro}

As a system undergoes spontaneous symmetry breaking, it
relaxes into a new phase. If the topology of the vacuum
manifold is non-trivial, there can be topological defects
that slow down the relaxation process. Eventually, however,
it is believed that the system will get to its ground state.
It is then somewhat surprising to find that certain systems
will essentially never get to the ground state. Instead they
will be trapped in a highly non-trivial state which, in one 
compact spatial dimension, consists of a lattice of domain walls.
                                                                                
In the present paper, we build upon a large body of earlier
work. Analysis of how simple domain wall networks 
relax may be found in 
Refs.~\cite{RydPreSpe90,KubIsiNam90,Kub92,GarHin03}. The kind of 
systems we are interested in lead to more complex domain walls
and are discussed in 
Refs.~\cite{PogVac00,Vac01,PogVac01,Pog02,Davetal02,Ton02,PogVac03,
Vac03,AntPogVac04}. Most of the discussion so far has been in one 
spatial dimension. Here we study a phase transition in two dimensions 
in which a network of domain walls is produced. Then we study the 
evolution of the network toward its final lattice state. We focus
on the likelihood that the system does not reach the vacuum, and
on the properties of the final lattice state.
     
\section{The model} 
\label{model} 
 
We start with the $SU(5)\times Z_2$ model:
\begin{equation} 
L = {\rm Tr} (\partial_\mu \Phi )^2 - V(\Phi ) 
\label{lagrangian} 
\end{equation} 
where $\Phi$ is an adjoint scalar field represented by 
$5\times 5$ Hermitian, traceless matrices. The potential
is assumed to satisfy $V(\Phi) = V(-\Phi)$ and the $Z_2$
symmetry is under $\Phi \rightarrow -\Phi$. If we truncate 
the model to just the four diagonal matrices, the only 
surviving symmetry is the permutation of the diagonal 
elements of $\Phi$. Now the model contains four real 
scalar fields $f_i$ ($i=1,..,4$), and the Lagrangian
is \cite{PogVac03}:
\begin{equation}
L = {1\over 2}\sum_{i=1}^4 (\partial_\mu f_i) ^2
     + V(f_1,f_2,f_3,f_4)
\label{s5lagrangian}
\end{equation}
We choose $V$ to be quartic in $\Phi$ leading to:
\begin{eqnarray}
V &=&
-{m^2\over 2}\sum_{i=1}^4 f_i^2
+ {h \over 4}(\sum_{i=1}^4 f_i^2)^2
+ {\lambda \over 8} \sum_{a=1}^3 f_a^4
\nonumber \\
&+& {\lambda \over 4} \left[
{7\over 30} f_4^4 + f_1^2 f_2^2 \right]
+ {\lambda \over {20}} [4(f_1^2 + f_2^2) + 9f_3^2] f_4^2
\nonumber \\
&+& {\lambda \over \sqrt{5}}f_2 f_4 \left( f_1^2
- {f_2^2 \over 3} \right) + {m^2 \over 4} \eta^2
\label{potential}
\end{eqnarray}
where the fields $f_i$ are defined by:
\begin{equation} 
\Phi(x)=f_1(x)\lambda_3+f_2(x)\lambda_8+f_3(x)\tau_3+f_4(x)Y \ , 
\label{components} 
\end{equation} 
where $\lambda_3$, $\lambda_8$, $\tau_3$ and $Y$ are the 
diagonal generators of $SU(5)$: 
\begin{eqnarray} 
\lambda_3&=&\frac{1}{2} {\rm diag}(1,-1,0,0,0) \ , \nonumber \\ 
\lambda_8&=&\frac{1}{2\sqrt{3}} {\rm diag}(1,1,-2,0,0), 
\nonumber \\ 
\tau_3&=&\frac{1}{2} {\rm diag}(0,0,0,1,-1) \ , \nonumber \\ 
Y&=&\frac{1}{2\sqrt{15}} {\rm diag}(2,2,2,-3,-3) \ . 
\label{ymatrix} 
\end{eqnarray} 
The model now has an $S_5 \times Z_2$ symmetry corresponding to
permutations of the 5 diagonal elements of the $SU(5)$ and the 
$Z_2$ transformation corresponding to change of sign. The potential 
is minimized by a vacuum expectation value of $f_4$ and the symmetry 
is broken down to $S_3\times S_2$.

The symmetry breaking:
\begin{equation}
S_5 \times Z_2 \rightarrow S_3\times S_2
\end{equation}
leads to $(5!/3!2!)\times 2 = 20$ distinct degenerate vacua. Each 
vacua is labeled by the vacuum expectation value of $\Phi$
and these will be denoted by:
\begin{eqnarray}
\pm (2,2,2,-3,-3)&,& \ \ \pm (2,2,-3,2,-3), \nonumber \\ 
\pm (2,-3,2,2,-3)&,& \ \ \pm (-3,2,2,2,-3), \nonumber \\  
\pm (2,2,-3,-3,2)&,& \ \ \pm (2,-3,2,-3,2), \nonumber \\ 
\pm (-3,2,2,-3,2)&,& \ \ \pm (2,-3,-3,2,2), \nonumber \\ 
\pm (-3,2,-3,2,2)&,& \ \ \pm (-3,-3,2,2,2)     \nonumber
\end{eqnarray}

\begin{figure}
\scalebox{0.50}{\includegraphics{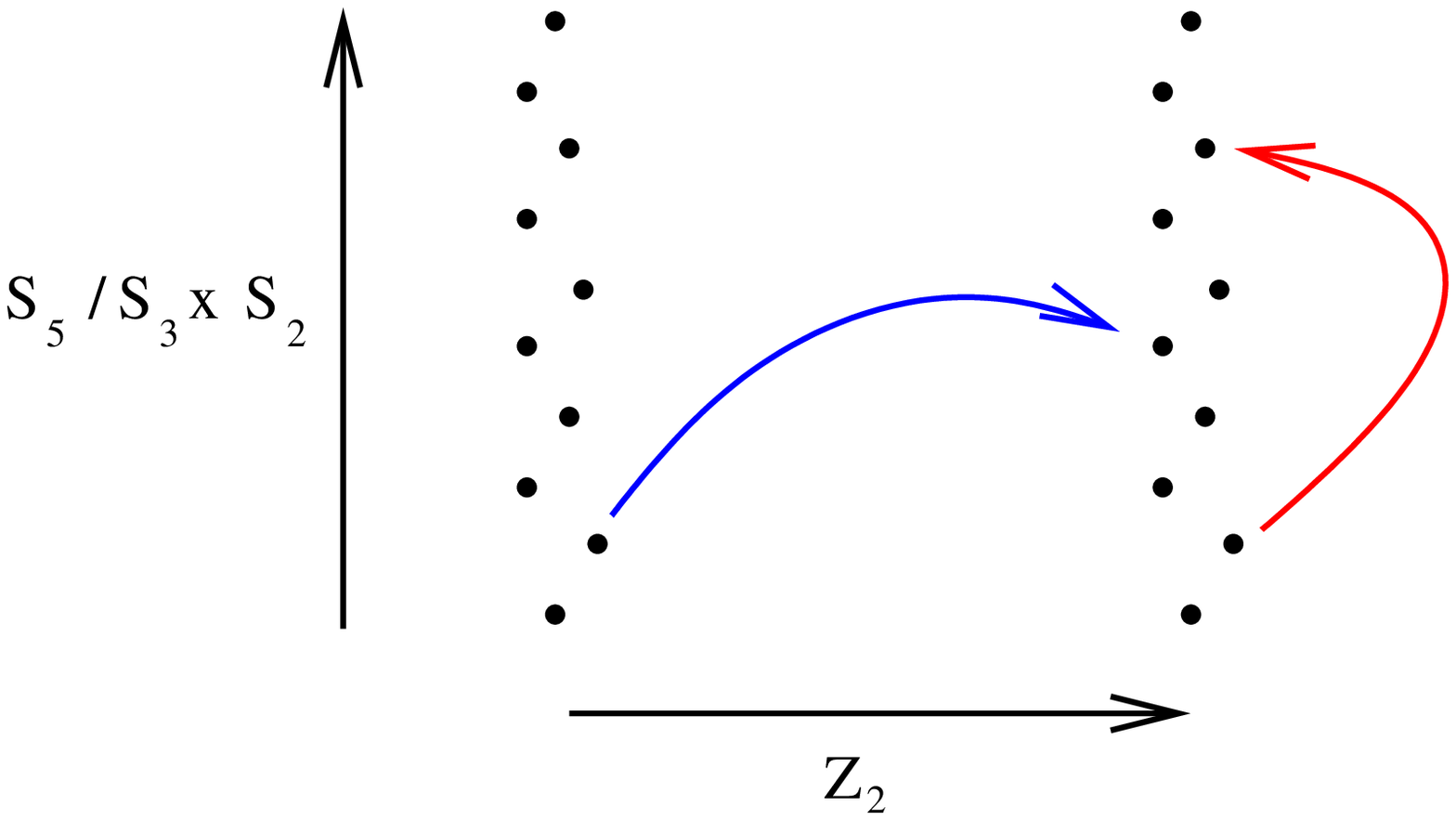}}
\caption{The vacuum manifold of the model consists of $10+10$ points. 
Domain walls correspond to solutions interpolating between any two 
vacua. There are two kinds of domain walls, one interpolating between 
the two sets of 10 vacua related by $Z_2$ transformations, and the 
other interpolating between vacua within a set of 10 vacua. If we
fix the vacuum at $x=-\infty$ there are 10 ways of choosing the
vacuum at $x=+\infty$ that differs by a $Z_2$ transformation,
leading to 10 different $Z_2$ walls. Of these 10 $Z_2$ walls,
3 are of lowest energy, 6 are of intermediate energy, and 1 
has the highest energy \cite{PogVac01}.
}
\label{fig1}
\end{figure}

We can find a domain wall solution interpolating between any two 
distinct vacua. Hence there are a large number of domain walls in 
the model. But there are two distinct
sets of domain walls (Fig~\ref{fig1}): the first set consists of 
domain walls across
which the vacua differ by the $Z_2$ transformation, while the second
consists of domain walls that separate vacua that do not differ
by $Z_2$. The 10 different $Z_2$ walls can further be distinguished 
according to their masses~\cite{PogVac01}. There are 3 
solutions that are least massive. As
an example, one of these lightest walls interpolates between 
vacua in the directions $(2,2,2,-3,-3)$ and $-(-3,-3,2,2,2)$ {\it i.e.}
vacua related by 2 permutations and a sign.
Then there are 6 intermediate mass walls between vacua related
by 1 permutation and a sign {\it e.g.} $+(2,2,2,-3,-3)$ and 
$-(-3,2,2,2,-3)$, and 1 heaviest wall interpolating between 
vacua related by 0 permutations and a sign {\it e.g.} 
$+(2,2,2,-3,-3)$ and $-(2,2,2,-3,-3)$. 

The charge of a wall (up to normalization) is defined by the
difference $\Phi(+\infty)-\Phi(-\infty)$. For the lightest $Z_2$ walls,
the charges are: $\pm (-1,-1,4,-1,-1)$ and permutations of the
entries. Hence even the lightest $Z_2$ wall comes in 5 different 
charges (position of the 4 entry) and there are 5 corresponding 
antiwalls. We shall say that a wall and antiwall are of the
``same type'' if their charges only differ by a minus sign.
Otherwise the walls are of ``different type''.

\begin{figure}
\scalebox{0.50}{\includegraphics{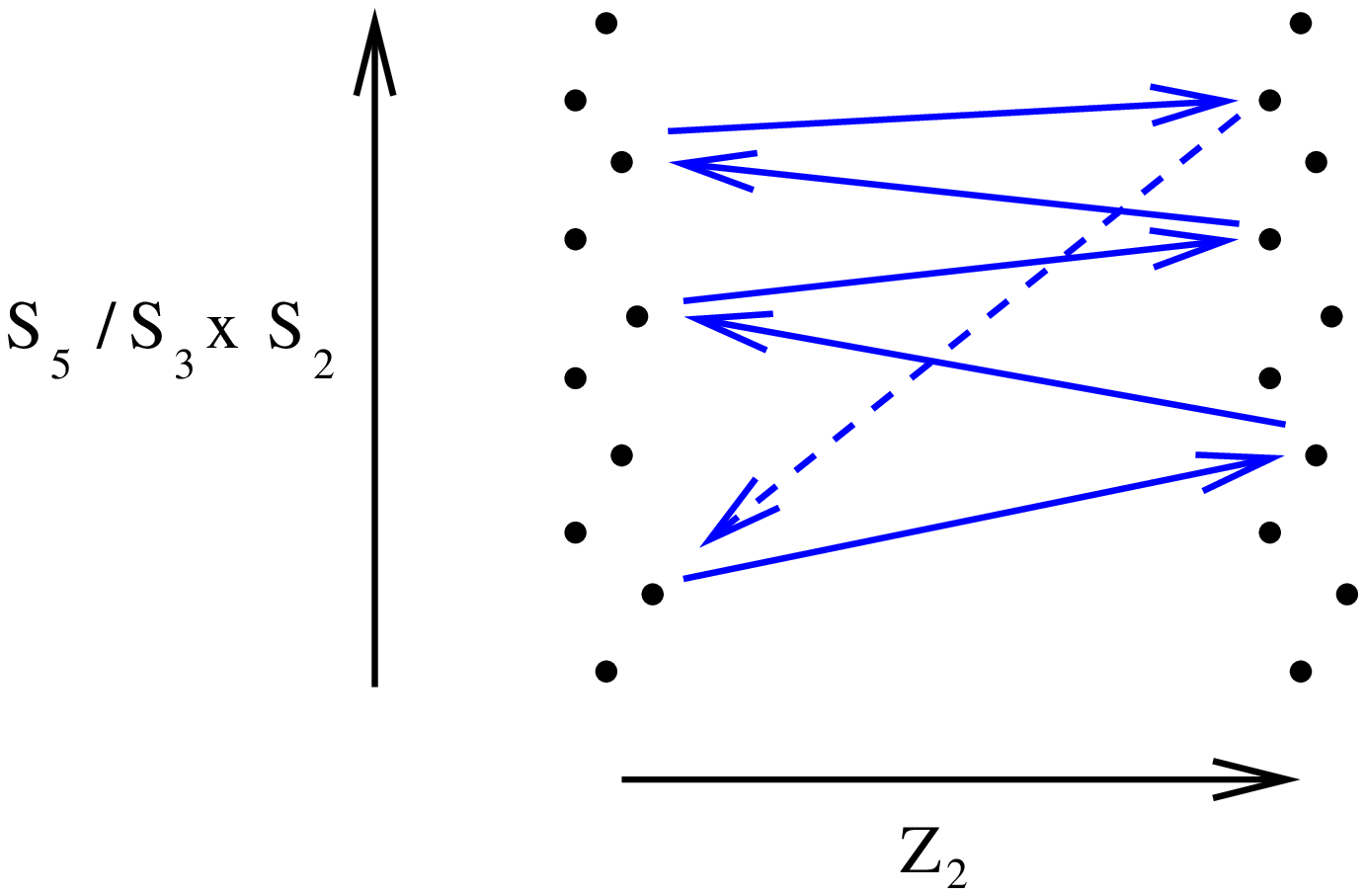}}
\caption{Depiction of the sequence of vacuum configurations in
the domain wall lattice.} 
\label{fig2}
\end{figure}

In Ref.~\cite{PogVac03} it was shown that walls and antiwalls
of different type repel in this model. This led to the possibility of
a stable lattice of alternating domain walls and antiwalls provided
that all the neighboring walls and antiwalls are of different type.
If a neighboring wall and antiwall are of the same type, they
will attract, then annihilate.
The lattice can be schematically depicted 
as in Fig.~\ref{fig2} or by the sequence of vacuum configurations:
\begin{eqnarray} 
... &\rightarrow& +(2,2,2,-3,-3) 
              \rightarrow -(2,-3,-3,2,2) \nonumber \\ 
              &\rightarrow& +(-3,2,2,-3,2) 
              \rightarrow -(2,-3,2,2,-3) \nonumber \\ 
              &\rightarrow& +(2,2,-3,-3,2) 
              \rightarrow -(-3,-3,2,2,2) \nonumber \\ 
              &\rightarrow& +(2,2,2,-3,-3) 
                                 \rightarrow ... 
\label{Higgs sequence} 
\end{eqnarray} 
This sequence gives the minimal lattice of 6 kinks. 

In our one dimensional simulations \cite{AntPogVac04}
we found that a domain wall lattice is inevitably formed after
a phase transition provided the compact dimension is not too small. 
In our preliminary work in two {\em equal} compact spatial dimensions 
($L_y=L_x$) we
had found that a domain wall network with junctions is formed after
the phase transition and this network coarsens, eventually ending
in the vacuum. In other words, a domain wall lattice was not
seen to form. The present paper deals with the case of two {\em unequal}
compact spatial dimensions ($L_y \ne L_x$). We expect that when one 
of the dimensions is very small, the system will behave like the
one dimensional case, and a lattice will almost certainly be 
formed. When the sizes of the two dimensions are comparable, based
on the preliminary results, we would not expect a lattice to form. 
This will turn out to be incorrect and our more extensive analysis 
will show that even in the $L_y =L_x$ case there is a small chance 
of lattice formation. We will be interested in determining the domain 
wall lattice characteristics as a function of the relative sizes
of the dimensions of the compact space.

\section{Numerical technique}
\label{numericaltechnique}

In our simulations we try to mimic some features of a typical
non-equilibrium second order phase transition. In order to
do that we use a Langevin-type equation based on the fundamental
Lagrangian of the theory Eq.~(\ref{s5lagrangian}):
\begin{equation} 
\left( \partial_t^2 -\nabla^2 \right) f_i + \partial_i V 
 + D\, \partial_t f_i = \Gamma_i,  
\label{eom} 
\end{equation} 
The Langevin equation describes the evolution of the system
coupled to a heat bath with temperature $T$. The effect of the 
bath is modeled by including in the equations of motion a
dissipation term with dissipation constant $D$, and a stochastic 
force $\Gamma_i(x,t)$. The stochastic force is a Gaussian distributed
field obeying: 
\begin{eqnarray} 
&& \langle \Gamma_i(x,t) \rangle = 0, \nonumber \\  
&& \langle \Gamma_i(x,t) \Gamma_j(x',t')\rangle= 2 D T \, \delta_{ij}  
\delta (x-x')\delta (t-t')  
\label{noise} 
\end{eqnarray} 
The amplitude of the noise in Eq.~(\ref{noise}) is chosen so as to  
guarantee that independently of the initial field configuration and 
of the particular value of the dissipation, the system will always 
equilibrate toward a thermal distribution with temperature $T$. 

In a typical simulation, we start with an arbitrary initial condition 
and evolve Eq.~(\ref{eom}) for a period of time long enough for the system
to equilibrate. We then mimic an instantaneous phase transition to $T=0$
temperature, by setting the noise term in Eq.~(\ref{eom}) to zero. In the
subsequent period of evolution, the dissipation forces the fields to its
possible vacua values, and a complex network of domain walls forms, 
separating different vacuum regions. This network is then allowed to evolve
 until a final stable state is reached. A detailed discussion on how
the evolution proceeds can be found in \cite{AntPogVac04}.

The equations of motion were discretized using a standard leapfrog 
method and periodic boundary conditions were used. The model 
parameters  were set to $m=1/(2 \sqrt{6})$, $\lambda=1/2$ and 
$h=-3/40$. For this particular choice of parameters there is a known
explicit analytical solution for the relevant domain profile \cite{PogVac00,Vac01}.
The lattice spacing was $\delta x=1.$ and the time-step 
set to $\delta t=.5$. For the parameters above, the wall core is resolved 
by more that 10 lattice points which should be accurate enough for 
the desired purposes. The value of the dissipation coefficient $D$  
does not influence the results during the stochastic stage 
of the simulation and we set it to $D=.5$ to ensure rapid thermalization.
We keep the same dissipation value during the evolution regime at $T=0$,
the reasons behind this choice and its effects on the final results are
discussed in detail in the next Section.

In order to identify the walls during and at the end of the simulation,
we convert the $f_i(x)$'s back into their original $SU(5)\times Z_2$ 
matrix form, $\Phi(x)$, via Eq.~(\ref{components}). It is easy to see 
that for the lowest energy kinks in the model, the sign of 
${\rm Tr}[\Phi^3]$ vanishes at the defect core \cite{AntPogVac04}. 
For these kinks, one (and only one) of the fields $\Phi_{jj}$ crosses 
zero at the core as well. In the simulation we measure, at each 
time-step, the value of ${\rm Tr}[\Phi^3]$ and look for pairs of 
neighboring lattice points for which this quantity changes sign. In
these cases we determine which of the field components goes through
zero. This way we are able both to find the domain walls and to 
identify them according to their five possible charges, as described 
in Section \ref{model}. 

All the simulations were run in two-dimensional domains with periodic
boundary conditions. For each set of runs, the size of the simulation
box in the $x$-direction was fixed to a constant $L_x$, whereas its 
$y$ dimension, $L_y$  was allowed to vary. 
For each choice of $L_x$ and $L_y$, we executed several independent
runs with different random seeds and averaged the relevant
physical quantities over those to obtain the final results. 
 
\section{Results}
\label{results}
 
\begin{figure}
\scalebox{0.50}{\includegraphics{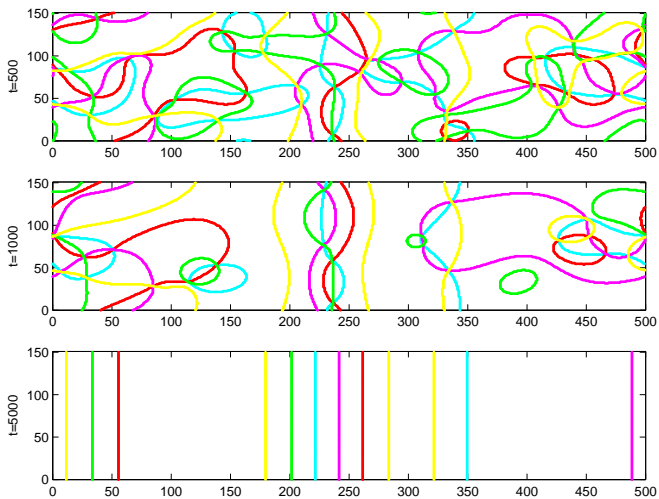}}
\caption{The domain wall network for three typical stages in an evolution 
in a toroidal domain, with dimensions $L_x=500$ and $L_y=150$. The different
shades correspond to the $5$ possible charges the domain walls can have in
the theory. Note that in the bottom figure there is a pair of neighboring
wall and anti-wall of the same type (the walls just before and after
the $300$ mark). These will later on annihilate each other leading to 
a final stable lattice consisting of $10$ walls.}
\label{fig3}
\end{figure}

In Fig.~\ref{fig3} we have an example of how the evolution of a typical
network of walls proceeds in a toroidal domain. In this case, the larger
dimension was set to $L_x=500$ and the smaller to $L_y=150$. For early 
times after the temperature is quenched to zero (top plot, $t=500$), we 
can observe a dense population of walls forming a complex network. 
While it is not obvious from the topology of the vacuum manifold
that domain wall junctions (``nodes'') exist in the model 
\cite{Davetal02}, 
the simulation clearly shows their presence. These nodes may be viewed 
as being due to the existence of domain wall lattices in one dimension. 
To see this, imagine a circle in two spatial
dimensions. It is possible to have a lattice of walls on this circle.
Now if we shrink the circle to a point, the walls forming the lattice
have to converge to a point. In other words, the walls on one of
the circles are like spokes on a wheel and have to necessarily have a 
point of convergence in two dimensions. While an infinite number of lattice 
configurations can
be constructed, the minimal number of walls in a stable lattice
is six. In the simulation also we see that a majority of intersecting 
nodes consist of six incoming walls. As the evolution proceeds, several 
mechanisms play a role in decreasing the overall wall density. Walls and 
anti-walls of the same type annihilate each other and closed loops can 
be formed later 
decaying into radiation. The intersecting nodes are also dynamical and 
can disentangle when nodes and anti-nodes collide. A snapshot of the 
wall network for intermediate times can be seen in the middle plot in 
Fig.~\ref{fig3}. Finally, for long times, all the loops collapse and all 
nodes annihilate, leaving behind a set of straight parallel walls wrapped 
around the shortest dimension of the simulation torus. Depending on their 
charge, some of these walls may still annihilate each-other, and the final 
state of the evolution is either the vacuum or a lattice of equidistant 
repelling walls. In the bottom plot of Fig.~\ref{fig3} we can see that 
for $t=5000$ the system has relaxed into a stack of parallel walls.
Since we know how to identify the wall charge from the simulation data, 
once this stage of the evolution is reached it is straightforward to predict 
which walls will annihilate and how many (if any) will remain in the final 
lattice. In this particular case there is only one wall anti-wall pair 
that annihilates, resulting in a lattice of 10 walls.

Though the pattern of the evolution remains the same independently of 
the relative dimensions of the two-dimensional torus, its final outcome 
depends markedly on the values of $L_x$ and $L_y$. 
 
\begin{figure}
\scalebox{0.50}{\includegraphics{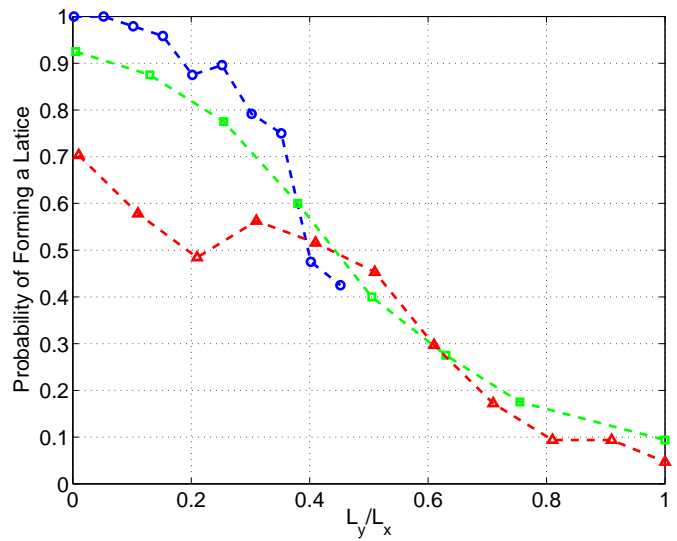}}
\caption{Probability of forming a lattice of domain walls as a function of 
the ratio of the two torus dimensions. $L_x$ is fixed for each curve, taking
the values $500$ (circles), $200$ (squares) and $100$ (triangles). Standard deviation
error bars (not shown for clarity) are of the same order of magnitude as the result.}
\label{fig4}
\end{figure}

The most striking trend in the data is that the final state of the evolution
is very likely to be a stable wall lattice when $L_y \ll L_x$, whereas the 
vacuum state is the predominant outcome when the two dimensions are comparable.
This dependence is illustrated more explicitly in Fig.~\ref{fig4},
where we show the probability of having a lattice forming as a function of the
ratio of the two dimensions of the simulation domain. For each choice of 
$L_x$ and $L_y$, the formation probability is defined as the number of 
independent runs that led to the formation of a lattice, divided by the total 
number of runs. For the three fixed choices of the largest dimension, 
$L_x=100$, $300$, and $500$, the results were obtained by averaging over 
$64$, $40$ and $40$ runs for each value of the smallest dimension $L_y$. 
In all three cases, $L_y$ was varied from $0$ to a larger value, close to 
$L_x$ in the two cases of small $L_x$ but only to about $250$ for $L_x=500$ 
(the reason being that for the larger case it takes a very long time for 
the system to reach its final state, and simulations for higher $L_y$ 
become numerically demanding). The errors calculated
from the standard deviation of the data are of the same magnitude as the
results. We chose not to show them in the plot (and likewise in the following
figures) so not to obscure the results.

For $L_y=0$, the probability of forming a lattice is quite high, being in fact
equal to unity in the case of $L_x=500$. This is to be expected since this case
corresponds to having the evolution taking place in a one-dimensional domain. 
As was shown in \cite{PogVac03}, in 1D, as the size of the physical domain 
goes to infinity, the probability of forming a lattice tends to one.
On the other hand, for the symmetric $2D$ system where $L_x=L_y$,
the lattice formation probability is quite small.
In the case of the $L_x=L_y=100$ simulations, out of $64$
independent runs only $3$ had a stable lattice of walls as their final state. 
A similar result was found for the $L_x=L_y=200$ case, where $3$ lattices 
were formed in a total of $32$ simulations. We also observed that in the
few cases where a lattice is formed in a $L_x=L_y$ box, the lattice shows
no preferred direction of alignment\footnote{In the $L_x=L_y$ case,
the domain wall lattice
spontaneously breaks the spatial symmetry under interchange of $x$ 
and $y$ in every member of the ensemble but the choice of direction 
by which it is broken is random.}. In the low and medium range of 
$L_y/L_x$ on the contrary, the walls in the final lattice are always parallel 
to the $y$ direction, as expected (see Fig.~\ref{fig3}). For $L_x=200$ the 
first lattices in the $x$ direction appear only for $L_y=150$ whereas for 
$L_x=100$ 
these are observed for values of the $y$ dimension larger than
$L_y=70$.

\begin{figure}
\scalebox{0.50}{\includegraphics{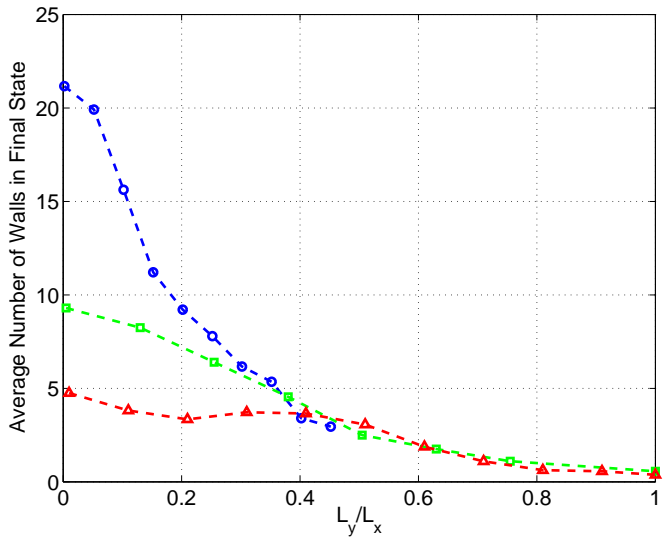}}
\caption{Average number of walls in the final stable configuration. The 
spatial dimensions of the torus are the same as in Fig.~\ref{fig4}.
Standard deviation
error bars (not shown for clarity) are of the same order of magnitude as the result.}
\label{fig5}
\end{figure}

It is interesting to note that the formation probability for large values of
the ratio $R\equiv L_y/L_x$ seems to be independent of the choice of $L_x$.
A similar effect can be seen in the averaged number of walls in the final
state, as plotted in Fig.~\ref{fig5}. The curves for the three choices of
$L_x$ converge for a  value of roughly $L_y/L_x=0.4$ and coincide above that.
Other quantities seem to follow the same pattern in this regime. The size ratio
above which walls parallel to the larger dimension are found in the final state,
is given by $R=150/200=0.75$ in the case $L_x=200$. For the smaller
box $L_x=100$, the result is similar with walls in the $x$ direction
appearing above $R=70/100=0.70$

\begin{figure}
\scalebox{0.50}{\includegraphics{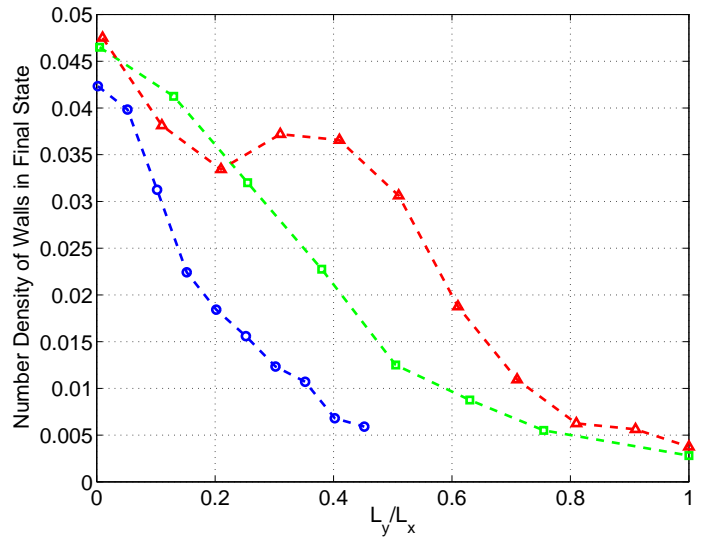}}
\caption{Total wall length per unit area  in the final stable configuration. 
The spatial dimensions of the torus are the same as in Fig.~\ref{fig4}.
Standard deviation error bars (not shown for clarity) are of the same 
order of magnitude as the result.}
\label{fig6}
\end{figure}

 The outcome of the evolution clearly seems to vary  within two well defined regimes.
For $R>0.4$ the properties of the final state depend only on the ratio of the
dimensions of the compact space, whereas for lower values of $R$ the results
depend also on $L_x$. In the small $R$ regime, for example,
the final number of walls varies considerably. That this is
to be expected can be easily seen by considering the $L_y=0$ limit, 
corresponding to the one-dimensional case.  In this situation we would
expect the final number of walls to be proportional to the length of the
1D domain.  In other words, the final wall density $\rho$ -- defined
as the total length in walls divided by the area of the lattice --
should depend little on  $L_x$, which is indeed observed: 
\begin{equation}
\rho_{100}=0.047,\,\,\,\,\,\,\,\rho_{200}=0.046,\,\,\,\,\,\,\,\rho_{500}=0.042
\end{equation}
As shown in Fig.~\ref{fig6} the curves for the number density are quite
close to each other for small values of $L_y/L_x$, diverging for higher
ratios of the two dimensions.

 Finally we note that, as mentioned in Section~\ref{numericaltechnique}, a non-zero
dissipation term is kept throughout the whole evolution. We expect that 
the presence of dissipation, though changing the type of scaling of the
dynamics of the network, should not influence qualitatively the outcome of the
evolution. The reason for this is that the probability of forming
a lattice should depend mostly on the number of uncorrelated domains in the
initial state of the evolution. As a check, we ran a few simulations
with dissipation set to zero during the evolution stage. The final 
results turned out to be different from the above, but only for small 
values of $L_x$ and $R$. The reason behind this is that after the quench 
there is still a lot of energy stored in the field which can be 
re-distributed among its different degrees of freedom. One consequence 
of this process is that it can effectively destroy the initial pattern of 
vacuum domains.  For example, in the $L_x=500$, $L_y=0$ case, the formation 
probability is still unity but the average wall number in the final state 
is less than shown in Fig.~\ref{fig4}. This observation can be interpreted 
as a consequence of the effective domain size being changed soon after the 
quench, when the dissipation is turned off.
This would also explain why the effects vanish for larger boxes. 
The conclusion is that keeping the dissipation on during the evolution allows us
to use smaller simulation boxes to probe the large scale limit of theory.

\section{Summary}
\label{summary}
 
We have numerically studied phase transitions in a model 
with $S_5\times Z_2$ symmetry breaking down to $S_3\times S_2$
in two compact spatial dimensions. As expected, the outcome
depends on the ratio of the sizes of the two dimensions:
$R \equiv L_y/L_x$. As $R \rightarrow 0$, the system behaves
as in one spatial dimension and a domain wall lattice
forms. As $R \rightarrow 1$, the system behaves as in
two spatial dimensions and a domain wall lattice rarely
forms. What is surprising though is that the critical
value of $R$ below which a lattice forms with large probability
is of order unity: $R_c \sim 0.4$.

In earlier work \cite{AntPogVac04} we have studied the evolution
of the network in time and found that the total length of the
network falls off as $t^{-0.7}$ in the $L_y=L_x$ ($R=1$) case. 
Here we have focused on the late time state of the system with
various values of $R$ and have determined the domain wall density 
$\rho$ at late times as a function of $R$. 
%
%
%
At the moment we do not know if there are naturally occurring
systems in which a domain wall lattice exists. The domain wall
lattice may also be relevant for cosmology, possibly in the context 
of brane cosmology.

\begin{acknowledgments} 
We are grateful to Levon Pogosian for collaboration on earlier
projects that led to this work. NDA was supported by PPARC. TV was supported
by DOE grant number 
DEFG0295ER40898 at Case. The numerical simulations were partially done 
on the COSMOS Origin2000 supercomputer supported by Silicon Graphics, 
HEFCE and PPARC. This work was partially supported by the
ESF COSLAB program.

\end{acknowledgments}

\end{document}